\documentclass[11pt,twoside]{article}

\usepackage{asp2014}

\aspSuppressVolSlug
\resetcounters

\begin{document}

\title{Planetary World Coordinate System in Astropy}

\author{Chiara~Marmo$^{1,}$$^3$ and St\'ephane~\'Erard$^2$}
\affil{$^1$oudeis.io, France; \email{marmochia.oudeis@proton.me}}
\affil{$^2$LESIA, Observatoire de Paris, Universit\'e PSL, CNRS, Sorbonne Universit\'e, Univ. Paris Diderot, Sorbonne Paris Cit\'e, Meudon, France}
\affil{$^3$Universit\'e Paris Saclay, Gif sur Yvette, France}

\paperauthor{Chiara~Marmo}{chiara.marmo@universite-paris-saclay.fr}{0000-0003-2843-6044}{Universit\'e Paris Saclay}{Facult\'e des Sciences}{Gif sur Yvette}{}{91190}{France}
\paperauthor{St\'ephane~\'Erard}{stephane.erard@obspm.fr}{0000-0002-9099-8366}{Observatoire de Paris}{LESIA}{Meudon}{}{92048}{France}


\begin{abstract}
Different research communities are involved in planetary coordinate standardization.
Geologists and Remote Sensing specialists work on extending Earth standards to Planets using
Geographical Information Systems (GIS) and coordinate descriptions endorsed by the Open
Geospatial Consortium (OGC).
Astronomers work to define Virtual Observatory (VO) metadata and FITS World Coordinate System (WCS)
for planetary bodies.
In this proceeding the implementation of the planetary WCS description in Astropy
is described. The related new features are available starting from Astropy 6.0.
The current work is part of a broader effort involving other consortia (e.g., heliophysics for Solar
or magnetospheric reference frames), and focuses on body-fixed frames to support surface and
atmosphere studies.
\end{abstract}
\section{Introduction}
Different research communities are interested and involved in planetary coordinate standardization.
Geologists and Remote Sensing specialists work on extending Earth standards to Planets using
Geographical Information Systems (GIS) and coordinate descriptions endorsed by the Open
Geospatial Consortium (OGC).
Astronomers work towards a definition of Virtual Observatory (VO) metadata and FITS World Coordinate
System for planetary bodies.
To improve interoperability between those worlds in \citet{marmoetal2018} we presented a
translation between GIS and FITS coordinate metadata, consistent with the Planetary Data System (PDS)
recommendations and with the planetary VO vocabulary as endorsed by the International
Virtual Observatory Alliance (IVOA) \citep{erardetal2022}.
Here we describe the implementation of that planetary WCS description in Astropy \citep{Astropy:2022}.

So far, Planetary WCS only consisted in accepted \verb"CTYPE" values possibly containing a two letter
code specific to each planet.
This is discussed in \citet{CalabrettaGreisen2002}, where data acquired by planetary satellites are
used as examples of application.
The 3-dimensional representation of a planetary body is currently missing, together
with the related standard keywords.
The keywords should describe the shape of the planetary object and the relative position between the observer and the planetary object.
Similar issues has been tackled in the past by the developers of Sunpy \citep{sunpy2020},
aiming to describe solar reference systems.
Unlike the Sun, planetary bodies can be significantly triaxial: this led to new developments
in astropy coordinate representations.
\section{Astropy developments}
\subsection{Planetary shapes}
A geodetic coordinate representation was available in Astropy since 2020, used to better transform to
and from Earth topocentric coordinate systems.
Its main attribute was a string definition of the standard spheroid needed to describe the Earth. 
Starting from Astropy 6.0 geodetic coordinate representations could be created as subclasses
of \verb"BaseGeodeticRepresentation" provided with the equatorial radius and flattening
assigned to \verb"_equatorial_radius" and \verb"_flattening" attributes.
Also, a new \verb"BaseBodycentricRepresentation" class has been implemented: it allows to instantiate
planetocentric coordinate representations via equatorial radius and flattening.
Using the representations above, any custom spheroid can be represented via planetodetic or
planetocentric coordinates.
As long as the coordinate frames share the same origin and orientation
this scheme makes straightforward the conversion between different
planetary representations, say the current bodycentric Mars description \citep{Seidelmannetal2002}
and the geodetic one defined in 1979 \citep{Daviesetal1980}, having also different flattening
(see Figure \ref{P615_1}).
\articlefigure[height=8cm]{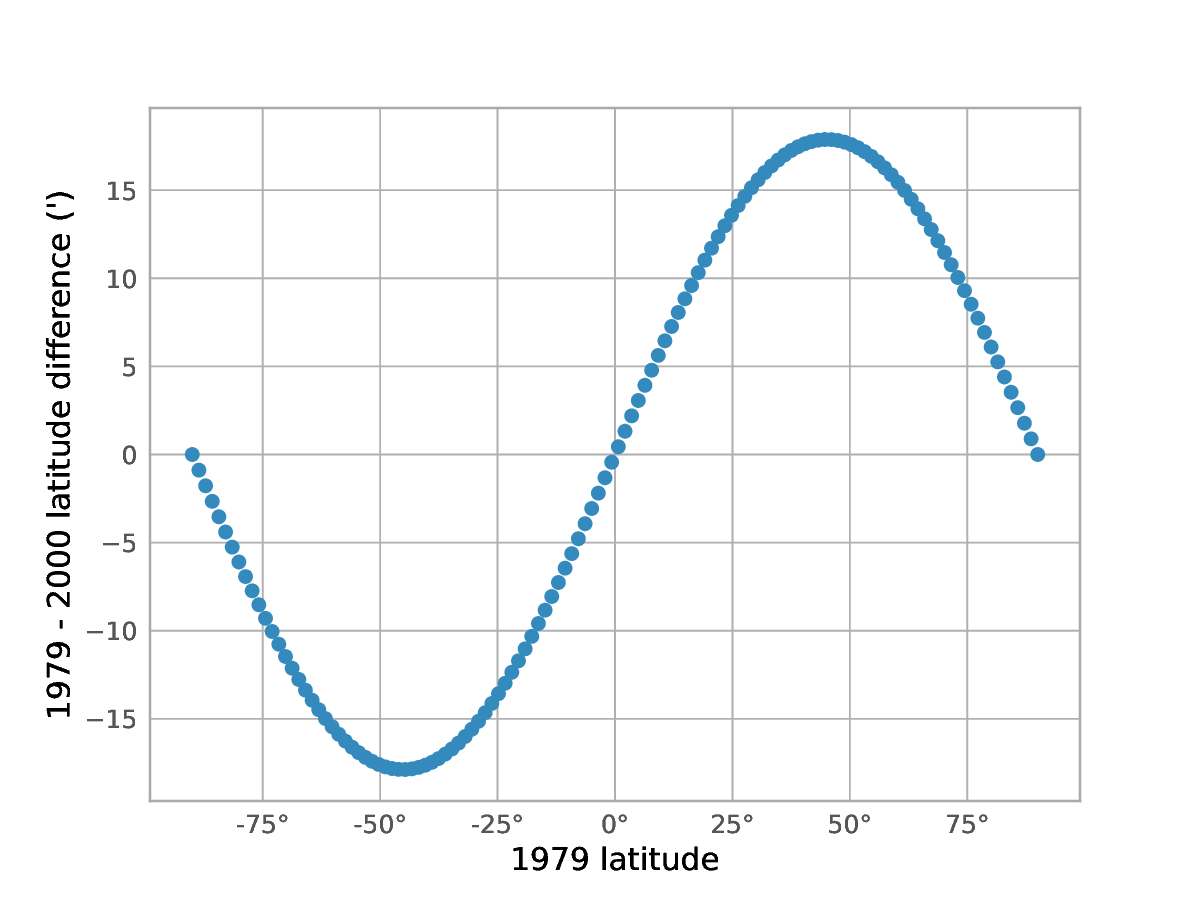}{P615_1}{Latitude differences between the 1979 planetodetic and
the 2000 planetocentric Mars description, in minutes.}
Standardized shape keywords allow to translate planetary WCS information in \verb"proj"
format\footnote{\url{https://proj.org/en/latest/usage/ellipsoids.html}}, leading
to automatic identification of the standard references defined in \citet{haremalapert2021}.
\subsection{Auxiliary WCS keywords}
Starting from version 6.0 Astropy ships a version of WCSlib greater than 8.0, extending the auxiliary
WCS structure to new planetary keywords (see table below).
\begin{table}[!ht]
\caption{Planetary FITS keywords included in the auxiliary WCS structure in WCSLib $>=$ 8.0}
\smallskip
\begin{center}
{\small
\begin{tabular}{ll}  
\tableline
\noalign{\smallskip}
Keyword & Description:\\
\noalign{\smallskip}
\tableline
\noalign{\smallskip}
A\_RADIUS & Semimajor axis of the ellipsoid: approximate shape used in projection. \\
B\_RADIUS & Intermediate axis of the ellipsoid: approximate shape used in projection. \\
C\_RADIUS & Semiminor axis of the ellipsoid: approximate shape used in projection. \\
BLON\_OBS & Bodycentric longitude of the observer in the body-fixed frame. \\
BLAT\_OBS & Bodycentric latitude of the observer in the body-fixed frame. \\
BDIS\_OBS & Distance between the centre of the celestial body and the observer.  \\
\noalign{\smallskip}
\tableline\
\end{tabular}
}
\end{center}
\end{table}
The three radii keywords translate the planetary shape into the FITS header.
The three observer's keywords prepare the path to the implementation of frame conversions.
Body-fixed planetary reference frame can be translated to and from a WCS description in FITS
headers.
Spatial coordinates can be mapped to pixel coordinates for image cut-outs or feature selection.
\section{Perspectives}
Body-fixed planetary frames are currently implemented as isolated frames: the next step would be
to add the transformations to astronomical frames (ICRS) and planetary topocentric frames
(planetary local horizontal coordinates).
This would support conversions, e.g., from landers and rovers coordinate frames to celestial ones,
as it is already possible in Astropy for Earth based observations.
\acknowledgements This work has been made possible thanks to the kind collaboration of the Astropy maintainers, in particular the main reviewer Marten van Kerkwijk.
The authors warmly thank Mark Calabretta, maintainer of WCSlib, for the keyword additions to WCSLib.

This work has been funded by the EuroplanetH2024 Research Infrastructure (RI) European project which has received funding from the European Union's Horizon 2020 research and innovation programme under grant agreement No 871149.
\bibliography{P615}  


\end{document}